\documentclass[twocolumn,showpacs,prl,superscriptaddress,floatfix]{revtex4}

\usepackage{color} 
\usepackage{tabularx} 
\usepackage{epsfig}
\usepackage{amsmath} 
\usepackage{amssymb} 
\usepackage{graphicx}
\usepackage{goodfloat}
\usepackage{times}

\begin{document}

\title{Density of States and Critical Behavior of the Coulomb Glass}

\author{Brigitte Surer}
\affiliation {Theoretische Physik, ETH Zurich, CH-8093 Zurich,
Switzerland}

\author{Helmut G.~Katzgraber} 
\affiliation {Theoretische Physik, ETH Zurich, CH-8093 Zurich,
Switzerland}
\affiliation{Department of Physics, Texas A\&M University, College Station,
             Texas 77843-4242, USA}

\author{Gergely T.~Zimanyi}
\affiliation {Department of Physics, University of California, Davis,
California 95616, USA}

\author{Brandon A.~Allgood}
\affiliation {Numerate Inc., 1150 Bayhill Drive, San Bruno, CA
94066, USA}

\author{Gianni Blatter}
\affiliation {Theoretische Physik, ETH Zurich, CH-8093 Zurich,
Switzerland}

\begin{abstract}

We present zero-temperature simulations for the single-particle density
of states of the Coulomb glass. Our results in three dimensions
are consistent with the Efros and Shklovskii prediction for the
density of states.  Finite-temperature Monte Carlo simulations
show no sign of a thermodynamic glass transition down to low
temperatures, in disagreement with mean-field theory. Furthermore, the
random-displacement formulation of the model undergoes a transition
into a distorted Wigner crystal for a surprisingly broad range of
the disorder strength.

\end{abstract}


\maketitle

The Coulomb glass (CG) is the earliest paradigm for understanding
the effects of strong disorder in electronic systems with long-ranged
interactions. Among its applications are the space-energy correlations
in transistors, the magnetization-switched metal-insulator transitions
in tunnel devices, the cotunneling magnetoresistance in ferromagnetic
devices, the ambipolar gate effect, the huge magnetoresistance
in semiconductor stacks, and the transparent refractory oxides,
to name a few.  In the CG, the Coulomb interaction remains long
ranged because the disorder localizes the electrons and thus impedes
screening. Therefore, the system forms its low energy states by
long-range configurational changes and avalanches.  After some early
approaches \cite{pollak:70,srinivasan:71}, Efros and Shklovskii
(ES) argued that the stability of the low-energy states against
the long-ranged single-particle dynamics requires the formation
of a soft ``Coulomb gap'' in the single-particle density of states
(DOS) of the form $\rho(E)\sim |E|^{\delta}$ with $\delta = D - 1$
as an upper bound and where the energy $E$ is measured from the Fermi
level \cite{efros:75}; $D$ being the space dimension. This Coulomb gap
leads to a typical variable range hopping form of the low-temperature
conductivity, i.e., $\sigma(T) = \sigma_0 \exp{[-(T_0/T)^{-1/2}]}$;
$\sigma_0$ and $T_0$ constant.

There is considerable experimental evidence in support of
these predictions from transport measurements of $\sigma(T)$
\cite{paalanen:86-ea,sarachik:02,yu:04,romero:05}, as well
as from tunneling conductance measurements of $\rho(E)$
\cite{massey:95,jaroszynski:96,butko:00-ea,grenet:07-ea}.  However,
subsequent theoretical considerations arrived at an exponential form
of the DOS by considering multi-electron ``polaronic'' processes:
$\rho(E) \sim \exp{[-(E_0/E)^{1/2}]}$ \cite{efros:76,pollak:92},
throwing the status of theoretical predictions for the DOS in question.

A large number of nonequilibrium glassy phenomena have been observed
in disordered electronic systems. These include slow dynamics
\cite{monroe:87-ea,benchorin:93,martinez-arizala:98-ea,ovadyahu:07},
aging, and memory effects \cite{vaknin:00-ea,orlyanchik:04,lebanon:05},
as well as changes in the noise spectrum \cite{bogdanovich:02}.
However, the existence of a thermodynamic glass transition cannot
be directly surmised from glassy dynamics. In fact, no well-defined
thermodynamic glass transition has been found in association with
these phenomena to date in three-dimensional (3D) systems.

These different theoretical predictions merged with the insight of
Pastor and Dobrosavljevic, who described a disordered electron system
with long-range interactions using the replica-based theoretical
framework of glass physics \cite{pastor:99,vojta:93}. Their work
offered a unified platform to analyze both the DOS and the glassy
characteristics of these systems. This programme was subsequently
expanded by the work of M\"uller, Ioffe and Pankov who included
replica symmetry breaking technology into their calculations
\cite{pankov:05,mueller:04,mueller:07}. All these studies concluded
that---within a mean-field approach---a soft Coulomb gap exists in
the single-particle DOS at $T=0$. Furthermore, for $T \le T_{\rm c}
\sim W^{-1/2}$, where $W$ is a measure of the disorder, the system
freezes into a ``Coulomb glass'' state. Note that the Coulomb glass
is analogous to a spin glass in a (random) field \cite{young:04}
which is known to not order.

The Coulomb glass has attracted considerable attention
numerically as well.  The initial work by Davies, Lee and Rice
reported the observation of a soft gap, but the data were
not conclusive with respect to the detailed functional form
of the DOS \cite{davies:82}.  Subsequent numerical studies
represented the disorder either by random site energies (CG)
\cite{baranovskii:79-ea,moebius:92,li:94,sarvestani:95-ea,wappler:97-ea,glatz:07}
or by random displacements (RD) between the sites
\cite{lee:88,grannan:93,diaz-sanches:00-ea,overlin:04,grempel:04,grempel:05}.
While the CG and RD models have different symmetries (and thus
possibly different universality classes \cite{vojta:94}), it
has nevertheless been argued that they both adequately capture
the key aspects of the real electronic system \cite{overlin:04}.
Different studies of the DOS in 3D have reported a DOS vanishing
at the Fermi level with $\rho(E) \sim |E|^\delta$ with $\delta =
2.1$ -- $2.6$ \cite{moebius:92,li:94}. Even more surprising was
the claim of a strongly disorder-dependent exponent $\delta$
\cite{overlin:04}.  Furthermore, studies attempting to locate a
transition to a glassy state were only successful in the RD model
\cite{grannan:93,overlin:04,grempel:04,grempel:05}.

The state of the field can be summarized as follows: A soft gap in the
DOS has been widely confirmed, but the predicted ES exponent $\delta =
D - 1$ is consistent only with experimental data, not with numerical
simulations. A true finite-temperature transition to a glassy state
has numerical support in the RD model but lacks evidence in the CG
model and in experiments.

Our results show that in the 3D CG model $\delta$ is close to $2$
and weakly disorder dependent \cite{mueller:07} and we find no
signature of a finite-temperature glass transition. Furthermore,
in the RD model the low-temperature ordering is indicative of a
distorted Wigner crystal.

\paragraph*{Model and numerical details.---}
\label{sec:model}

The Coulomb glass (CG) Hamiltonian is given by \cite{efros:75}: 
\begin{equation}
{\mathcal H}_{\rm CG}
= \frac{1}{2}\sum_{i \neq j}(n_i - \nu)\frac{e^2}{\kappa r_{ij}}(n_j - \nu) 
+ \sum_i n_i \varepsilon_i ,
\label{eq:ham_cg}
\end{equation}
where $n_i \in\{0,1\}$ is the electron number at site $i$, $\nu$ the
filling factor, and $e^2/\kappa r_{ij}$ the Coulomb repulsion. The
sites lie on a three-dimensional lattice of size $N = L^3$, and
the electron number is coupled to Gaussian-distributed random site
energies $\varepsilon_i$ with zero mean and standard deviation
$W$, i.e., ${\mathcal P}(\varepsilon_i) = (2\pi W^2)^{-1/2} \,
\exp(-\varepsilon_i^2/2W^2)$.  In the RD model, instead of random
site energies, the disorder is represented by Gaussian-distributed
random displacements of the lattice sites with standard deviation
$\sqrt{3}W$. The DOS is given by the disorder average of $\rho(E)
= (1/N)\sum_i\delta(E - E_i)$ with $E_i = \sum_{j \neq i}(n_j -
\nu)(e^2/\kappa r_{ij}) + \varepsilon_i$ the local single-particle
energy \cite{efros:75}.

For the simulations we use particle-conserving dynamics and periodic
boundary conditions. To cope with the long-range Coulomb interactions
we perform a resummation technique in which we sum all interactions
over periodic images and renormalize the energy scales such that the
nearest-neighbor distance is $a = 1$.  To compute the ground-state DOS
($T = 0$) we use extremal optimization \cite{boettcher:01}.  For the
CG model we perform $2^{19}N$ updates and study systems of up to $N =
14^3$ sites in 3D for $W = 0.2$ and $0.4$ and average over $3000$ disorder samples for $L \le 12$ and $1800$ ($800$) samples for $L =
14$ for $W=0.2$ ($W=0.4$). For the RD model we study $N =
14^3$ sites and average over $100$ disorder samples (fluctuations are
small).  For the study at finite temperatures we use exchange Monte
Carlo \cite{hukushima:96,comment:classic}.  Equilibration is tested
by a logarithmic data binning. Once the last three bins agree within
errors, the system is in thermal equilibrium.  Simulation parameters
can be found in Table \ref{tab:simparams}.

\begin{table}
\caption{
Top: Simulation parameters for the simulations of the 3D Coulomb glass
model with Gaussian disorder of strength $W$ at finite temperature.
$L$ is the system size, $N_{\rm sa}$ is the number of disorder samples,
$N_{\rm sw}$ is the number of equilibration sweeps, $T_{\rm min}$ is
the lowest temperature, $T_{\rm max} = 0.455$ the highest temperature
and $N_{\rm r}$ the number temperatures used in the exchange Monte
Carlo method.  Temperatures are measured in units of ${e^2}/{\kappa
a}$, $a = 1$ being the lattice constant. Bottom: Parameters for the
3D RD model simulations.
\label{tab:simparams}}
{\footnotesize
\begin{tabular*}{\columnwidth}{@{\extracolsep{\fill}} l@{\hspace{-2.25em}}r r r c c }
\hline
\hline
$W$ & $L$  &  $N_{\rm sa}$  & $N_{\rm sw}$ & $T_{\rm min}$ & $N_{r}$\\
\hline
$0.20$ & $6$  & $4\,290$ & $2^{18}$ & $0.030$ & $27$ \\
$0.20$ & $10$ &    $388$ & $2^{18}$ & $0.030$ & $27$ \\
$0.20$ & $14$ &    $251$ & $2^{18}$ & $0.083$ & $17$ \\
\hline
$0.40$ & $6$  & $4\,955$ & $2^{18}$ & $0.030$ & $27$ \\
$0.40$ & $10$ &    $148$ & $2^{18}$ & $0.030$ & $27$ \\
$0.40$ & $14$ &     $98$ & $2^{18}$ & $0.083$ & $17$ \\
\hline
\hline
$W$ & $L$  &  $N_{\rm sa}$  & $N_{\rm sw}$ & $T_{\rm min}$ & $N_{r}$\\
\hline
$0.10$ & $8$  &    $173$ & $2^{20}$ & $0.030$ & $27$ \\
$0.20$ & $8$  &    $133$ & $2^{20}$ & $0.030$ & $27$ \\
$0.40$ & $8$  &     $93$ & $2^{20}$ & $0.030$ & $27$ \\
$0.80$ & $8$  &    $143$ & $2^{20}$ & $0.030$ & $27$ \\
\hline
\hline
\end{tabular*}
}
\vspace*{-0.50cm}
\end{table}

\paragraph*{Results for the density of states.---}
\label{sec:dos}

\begin{figure}

\includegraphics[width=8.0cm]{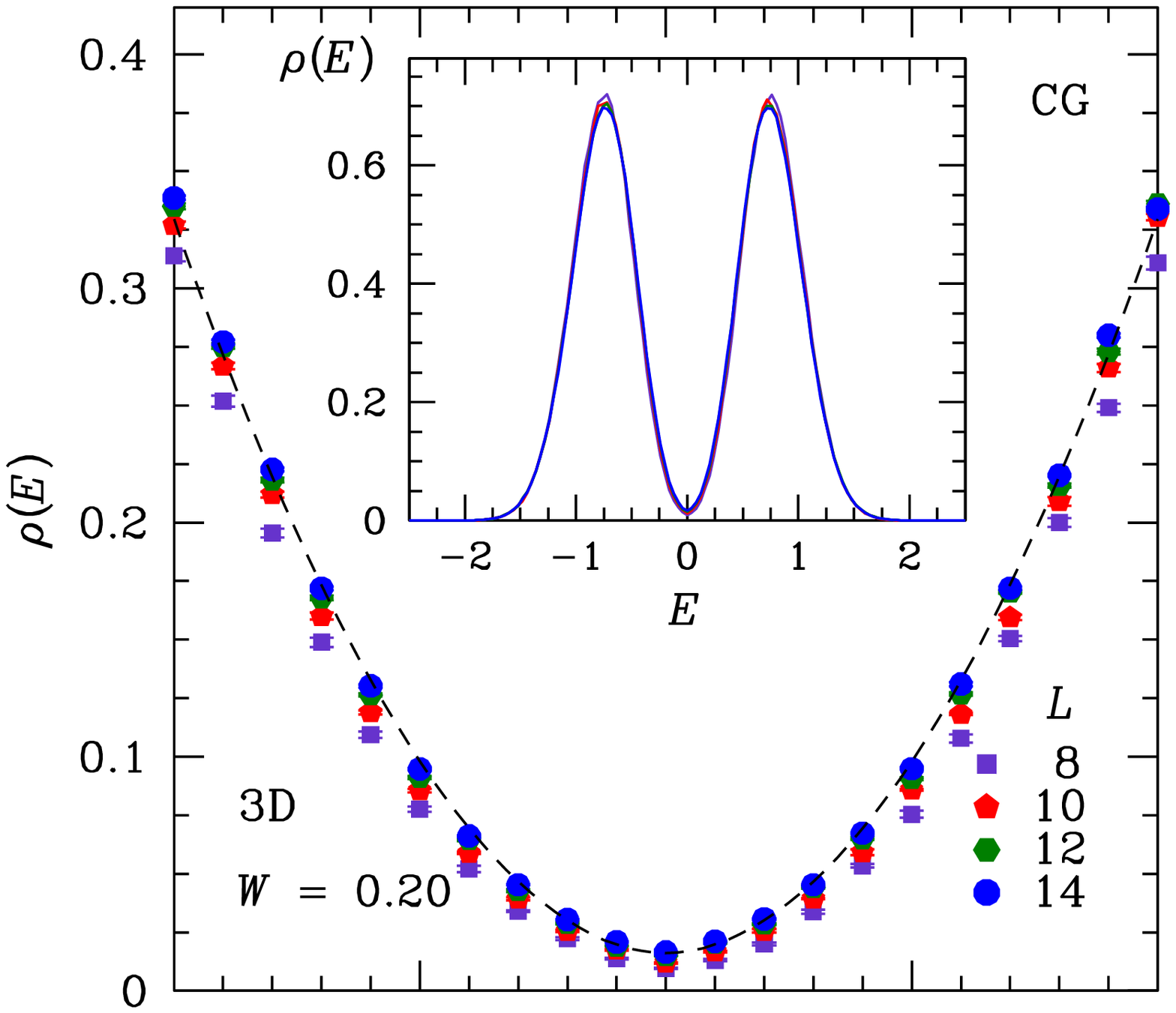}

\vspace*{-2.30cm}

\includegraphics[width=8.0cm]{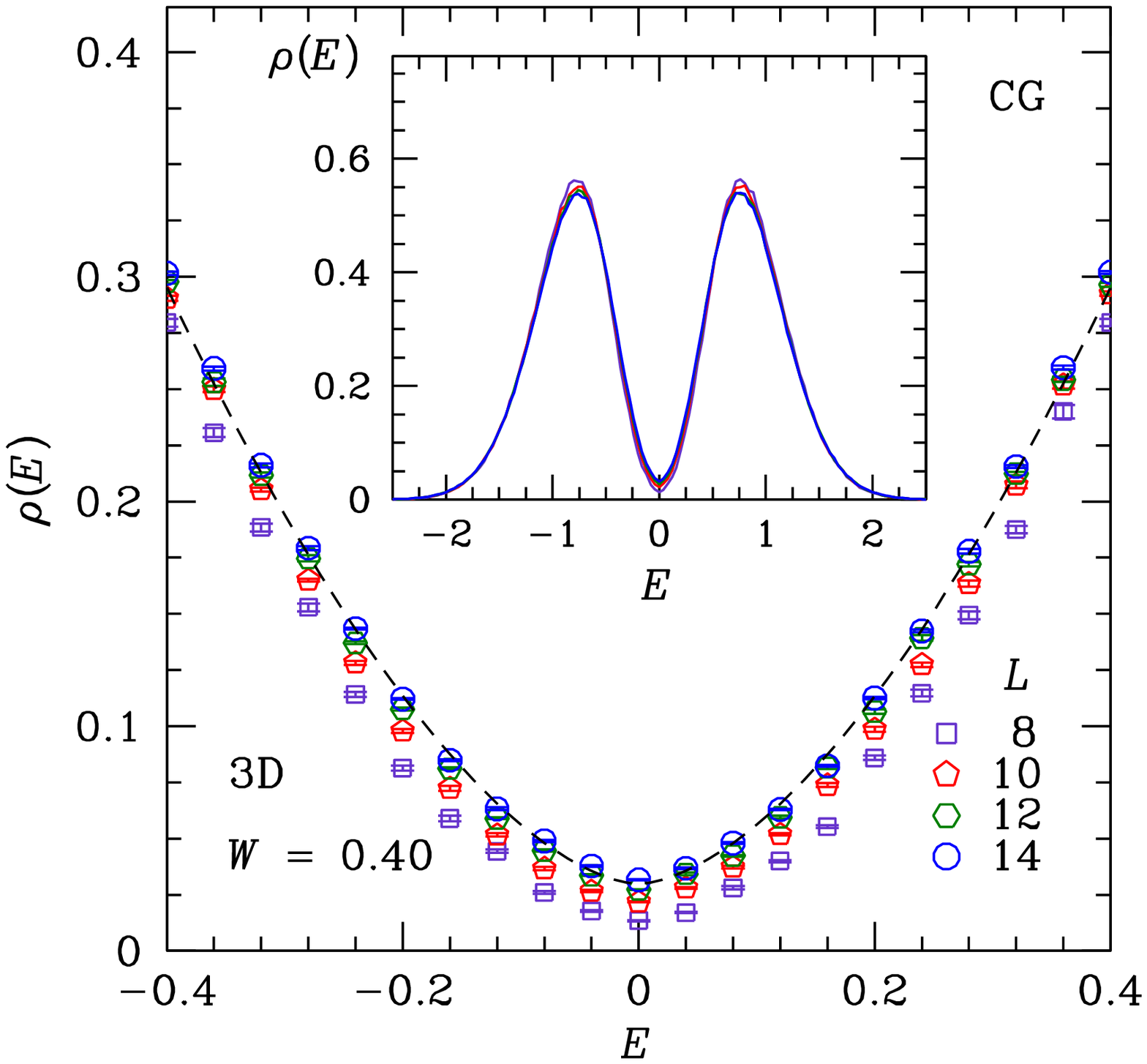}

\vspace*{-1.50cm}

\includegraphics[width=8.0cm]{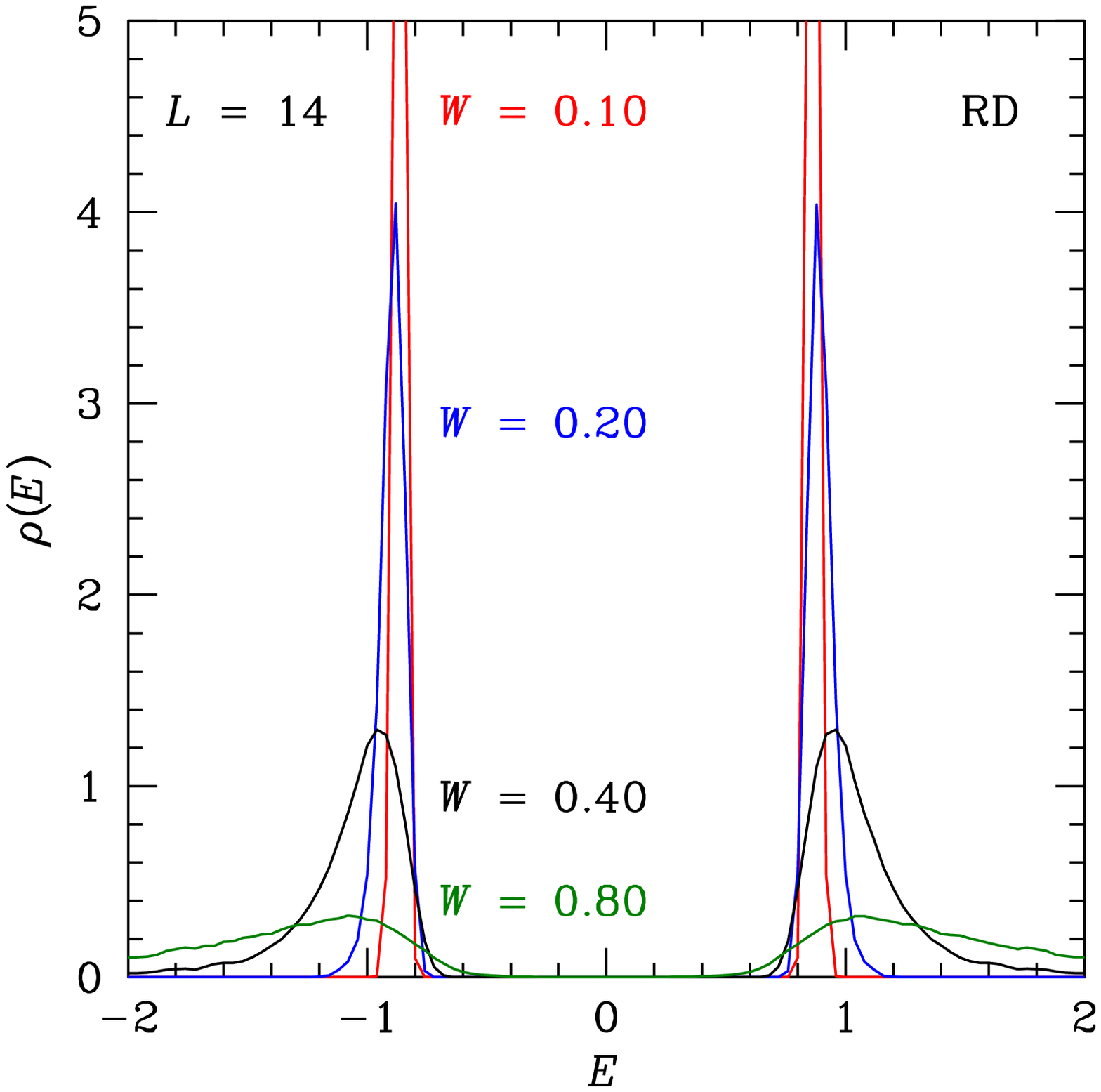}

\vspace*{-1.20cm}

\caption{(Color online)
Top: DOS for the 3D CG model for $W = 0.20$. The data are well
fit by $\rho(E)\sim E^\delta$, $\delta \approx 2$ (dashed lines
are a guide to the eye) around the Fermi level. Center: Same as
in the top panel for $W = 0.40$.  The insets show the full DOS.
Both panels have the
same horizontal range. Bottom: DOS for the 3D RD model.  For all $W$
studied the data show a bimodal structure with peaks at $|E| \sim 1$
and a hard gap of size $\sim 2$, in stark contrast to the CG model.
}
\label{fig:DOS}
\end{figure}

Figure \ref{fig:DOS} (top and center panels) show the DOS at $T = 0$
for the 3D CG model for two disorder strengths close to the Fermi level
($E = 0$) at half filling ($\nu = 1/2$); the insets show the whole
functional shape. The data can be fit very well with a form $\sim
|E|^\delta$ with $\delta = 2.01(2)$ $(L=14)$ for $W = 0.2$ 
and $\delta = 1.83(3)$ $(L=14)$ for $W = 0.4$ (restricted to $|E|\le 0.3$), which is
close to the ES value of $\delta
\approx D - 1$.

Fig.~\ref{fig:DOS} (bottom) shows the DOS of the RD model for $L =
14$. The DOS shows a pronounced double-peak, the width of the peaks
dependent on $W$. There is no sign of the characteristic Coulomb gap
shape, moreover the peaks at $|E|\sim 1$ are typical of a Wigner
crystal (WC). Thus the DOS of the RD model is indicative of the
formation of a moderately-distorted WC at $T = 0$.

\paragraph*{Results at finite temperature.---}
\label{sec:crit}

At half filling ($\nu = 1/2$) the ground state of the {\it clean}
system ($W = 0$) is a WC with a bipartite charge pattern.  For a WC
the DOS is expected to be two delta functions, separated by a charge
gap $E_{\rm WC}$.  The energy required to move a particle from a site
on the occupied sublattice to a site on the unoccupied sublattice is
$E_{\rm WC} \approx 2$ in units of ${e^2}/{\kappa a}$.  Since the
peaks of the DOS of the CG are approximately centered around $|E|\sim
1$ (Fig.~\ref{fig:DOS}, inset), it needs to be verified that the
observed DOS is indeed representative of a glassy phase and not only
that of a distorted WC.  Therefore we study the nature of the phase
at finite $T$ by computing both an order parameter for a glassy state,
\begin{equation}
q_{\rm GL} = 	\frac{4}{N} 
		\sum_{i = 1}^{N} (n_i^{\alpha}-1/2) (n_i^{\beta}-1/2) ,
\label{eq:opcg}
\end{equation}
and an order parameter for the competing Wigner crystal 
\begin{equation}
m_{\rm WC} = \frac{2}{N} \sum_{i = 1}^{N}(-1)^i (n_i-1/2) .
\label{eq:opwc}
\end{equation}
In Eq.~(\ref{eq:opcg}) $\alpha$ and $\beta$ refer to two copies
of the system with the same disorder \cite{comment:nu}.
If the system forms a Wigner crystal we expect $[\langle m_{\rm
WC}\rangle]_{\rm av} \to 1$ for $T \le T_c$, whereas if the system
freezes into a glass we expect $[\langle q_{\rm GL}\rangle]_{\rm av}
\to 1$ and $[\langle m_{\rm WC}\rangle]_{\rm av} \to 0$ for $T \to 0$.
Here $[\cdots]_{\rm av}$ denotes the average over disorder and $\langle
\cdots \rangle$ is a thermal average.

\begin{figure}

\includegraphics[width=8.0cm]{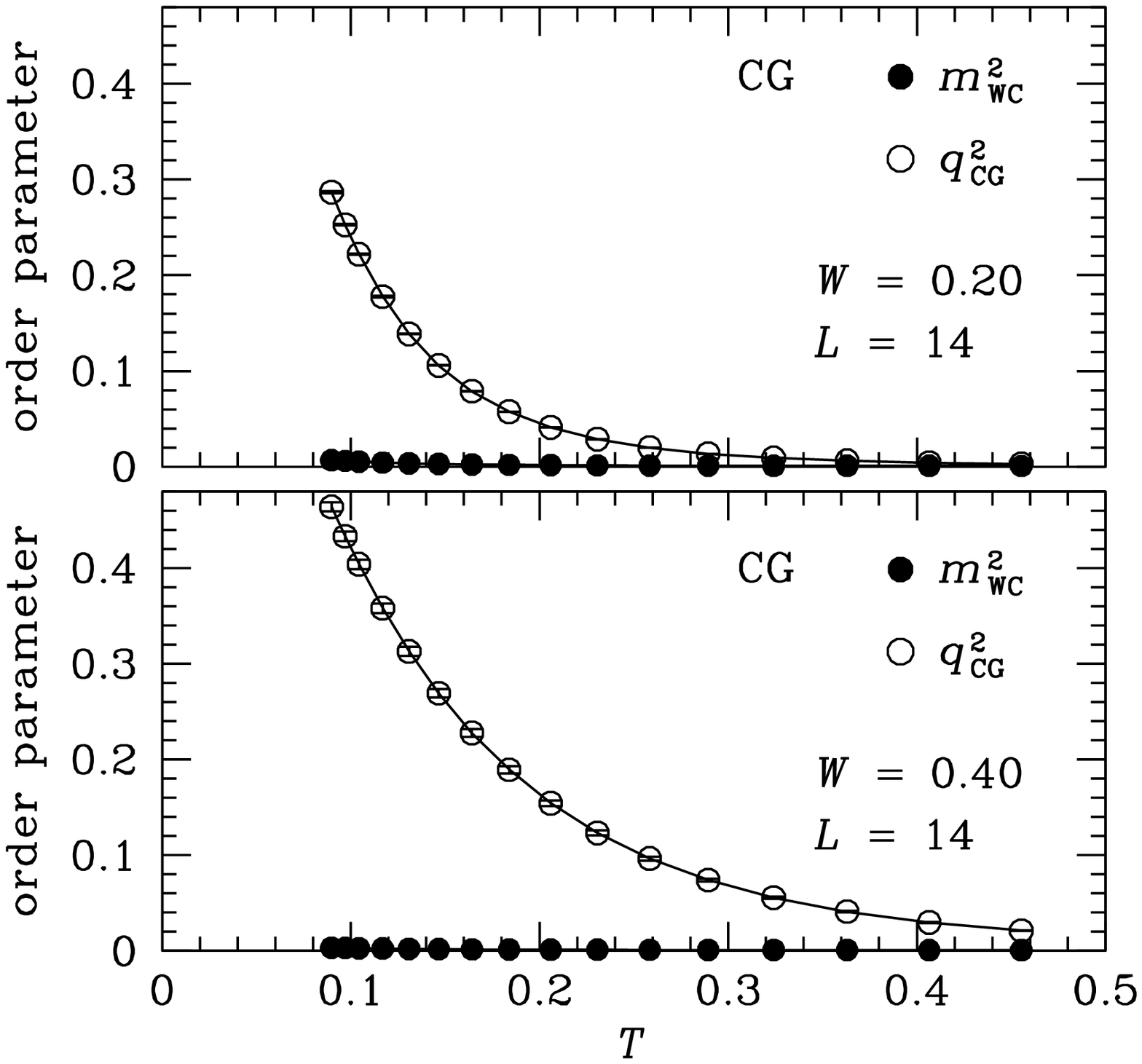}

\vspace*{-1.50cm}

\includegraphics[width=8.0cm]{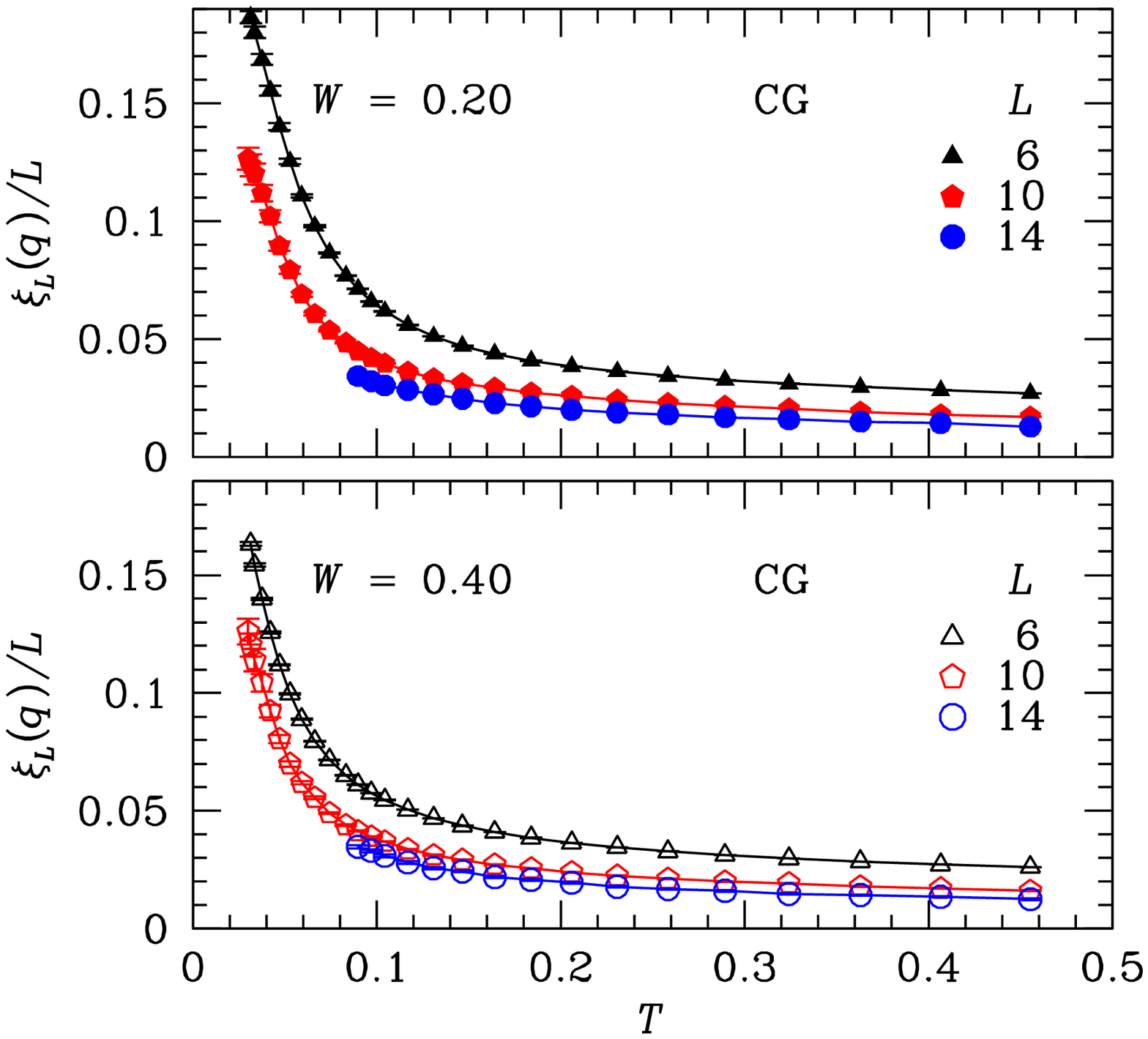}

\vspace*{-1.50cm}

\includegraphics[width=8.0cm]{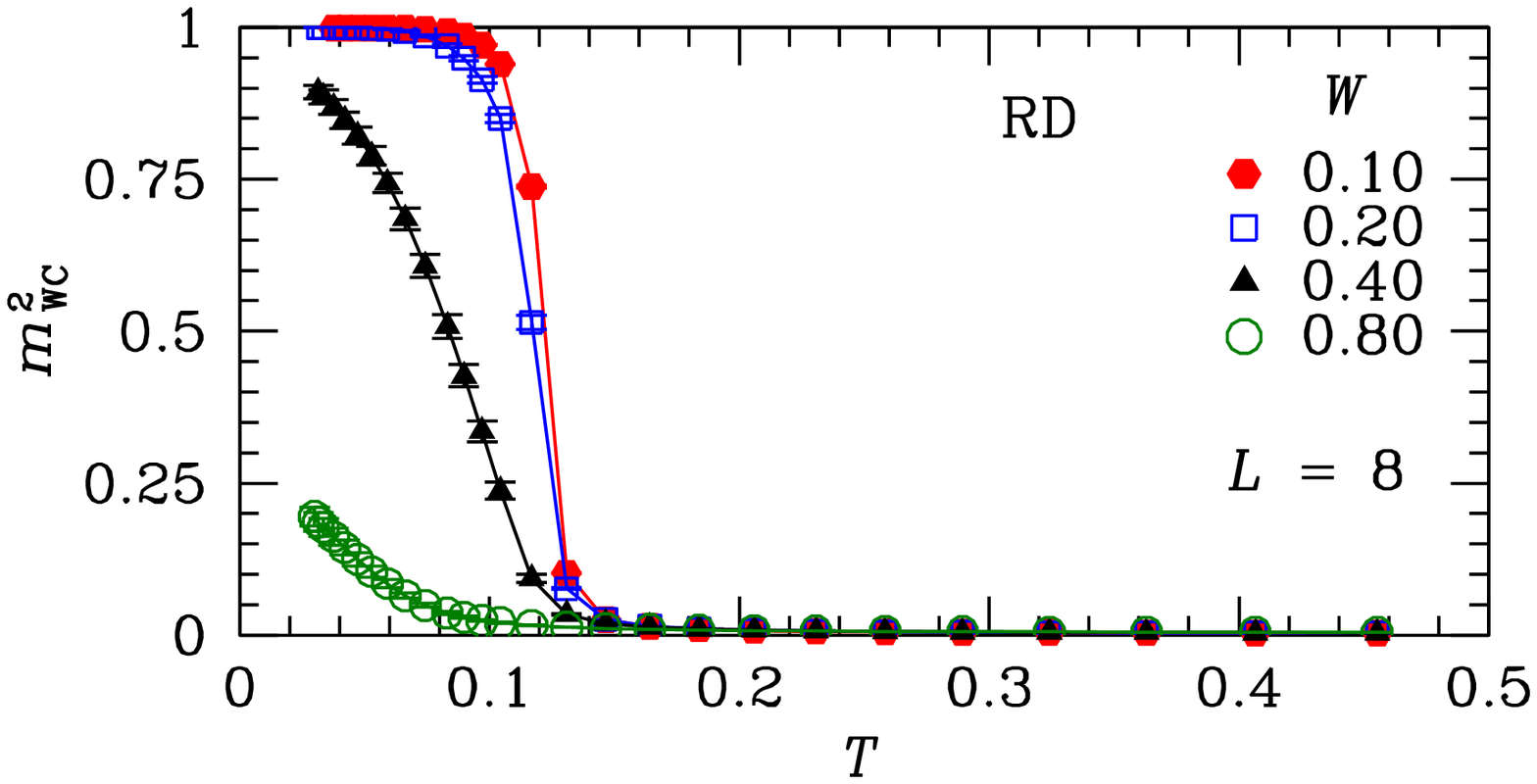}

\vspace*{-0.60cm}

\caption{(Color online)
Top: Wigner crystal order parameter $[\langle m^2_{\rm WC}\rangle]_{\rm
av}$ and glass order parameter $[\langle q^2_{\rm GL}\rangle]_{\rm av}$
as a function of temperature $T$ for different disorder strengths $W$
for the CG model. In all cases $[\langle m^2_{\rm WC}\rangle]_{\rm av}
\ll [\langle q^2_{\rm GL}\rangle ]_{\rm av}$. Center: Finite-size
correlation length as a function of $T$ for different disorder
strengths and system sizes for the CG model. The data show no
crossing, i.e., the absence of a thermodynamic transition for the
studied temperature range.  Bottom: Wigner crystal order parameter
for the RD model ($L = 8$) as a function of temperature for different
$W$. For $T \lesssim 0.1$, which quantitatively agrees with the
critical temperatures estimated in Ref.~\cite{overlin:04},
crystalline order emerges.
}
\label{fig:op}
\end{figure}

To locate the putative glass transition we compute the two-point
finite-size correlation length \cite{ballesteros:00-ea} given by
\begin{equation}
  \xi_{\rm GL}(L,T) = \frac{1}{2 \sin (|{\bf k}_\mathrm{min}|/2)}
  \left[\frac{\chi({\bf 0})}{\chi({\bf k}_\mathrm{min})} 
    - 1\right]^{1/2},
  \label{eq:xiL}
\end{equation}
where ${\bf k}_\mathrm{min} = (2\pi/L,0,0)$ is the smallest nonzero
wave vector and $\chi({\bf k})$ is the Fourier transform of the
susceptibility $\chi = [\langle q_{\rm GL}^2 \rangle - \langle q_{\rm
GL} \rangle^2]_{\rm av}$.  We use four replicas to compute $\chi$
to avoid biases. Because $\xi/L \sim X[L^{1/\nu}(T - T_{c})]$,
a phase transition at $T_c$ is signaled by the correlation lengths
for different $L$'s crossing at the same $T = T_c$.

Figure \ref{fig:op} (top panel) shows the $q^2_{\rm GL}(T)$ and
$m^2_{\rm WC}(T)$ order parameters as a function of temperature for
different disorder strengths in the CG model. The glass order parameter
increases as the $T \to 0$, whereas the Wigner crystal order parameter
does not exhibit any ordering tendency, $m_{\rm WC}(T)$ remaining $\sim
40$ ($140$) times smaller than $q_{\rm GL}(T)$ for $W = 0.2$ ($W =
0.4$) at $T = 0.08$.  Figure \ref{fig:op} (center panel) shows the
correlation length for the glass order parameter as a function of $T$
for the CG model. The data do not cross for the studied temperatures
and thus there is no sign of a transition for $T \ge 0.03$, disagreeing
with mean-field predictions \cite{mueller:04,mueller:07}. The lack
of a transition is mirrored by the small correlation length and the
proximity to the ground-state energy (not shown).

In Fig.~\ref{fig:op} (bottom panel) we show $m^2_{\rm WC}$ for the RD
model for disorder strengths up to $W = 0.8$ covering the disorder
range studied in Ref.~\cite{overlin:04}. For all $W$ studied,
$m^2_{\rm WC}$ rises noticeably (in contrast to the CG model).
This further underlines that---for the studied disorder range---the
phase transition in the RD model occurs into a surprisingly robust
distorted Wigner crystal phase.

\paragraph*{Conclusions.---}
\label{sec:conclusions}

We have analyzed the Coulomb glass at low and zero temperature
and find that the gap exponent of the density of states is close
to $\delta \approx D - 1$ in 3D systems.  Furthermore, we find
no evidence of a finite-temperature transition into a CG phase
in 3D for $W = 0.2$ and $0.4$. This suggests that the CG in 3D is
at or below its lower critical dimension, which would explain the
discrepancy with the mean-field results predicting a finite transition
temperature.  Finally, we have shown that in a broad disorder range
the random-displacement version of the CG model orders into a distorted
Wigner crystal and not into a glassy state.

\begin{acknowledgments} 

We thank S.~Boettcher, V.~Dobrosavljevic, E.~Grannan, N.~Jensen,
M.~M\"uller, M.~Palassini, S.~Pankov, D.~Popovic, V.~Oganesyan,
M.~Schechter, D.~Sherrington, B.~Shklovskii, T.~Vojta, and A.~P.~Young
for discussions.  The simulations have been done on the ETH Z\"urich
clusters. H.G.K.~acknowledges support from the SNF under Grant
No.~PP002-114713.

\end{acknowledgments}

\vspace*{-1.5em}

\bibliography{refs,comments}

\begin{thebibliography}{46}
\bibitem[{com({\natexlab{a}})}]{comment:classic}
\bibinfo{note}{It is enough to treat the problem classically since the
  electrons are localized by the disorder and $n_i \in \{0,1\}$, i.e., there is
  no need to take exclusion effects into account.}
\bibitem[{com({\natexlab{b}})}]{comment:nu}
\bibinfo{note}{All simulations are done with $\nu = 1/2$. Accordingly,
  Eqs.~(\ref{eq:opcg}) and (\ref{eq:opwc}) are defined for this special case.}
\end{thebibliography}

\end{document}